\documentclass[twocolumn,english,prl]{revtex4}
\usepackage[T1]{fontenc}
\usepackage[latin1]{inputenc}
\usepackage{color}
\usepackage{babel}
\usepackage{amsmath}
\usepackage{graphicx}
\usepackage{amssymb}
\usepackage{esint}
\usepackage{babel}

\begin{document}

\title{Analytical results from the quantum theory of a single-emitter nanolaser. II.}
\author{Nikolay V.~Larionov$^{1}$}
\email{larionov.nickolay@gmail.com}
\author{Mikhail I.~Kolobov$^{2}$}
\affiliation{$^{1}$Department of Theoretical Physics,
St-Petersburg State Polytechnic University, 195251,
St.-Petersburg, Russia}
\affiliation{$^{2}$Laboratoire de Physique
des Lasers, Atomes et Molécules, Université Lille 1, 59655
Villeneuve d'Ascq Cedex, France.}

\begin{abstract}
Using two equivalent approaches, Heisenberg-Langevin and density
operator, we investigate the properties of nanolaser: an
incoherently pumped single two-level system interacting with a
single-cavity mode of finite finesse. We show that in the case of
good-cavity regime the Heisenberg-Langevin approach provides the
analytical results for linewidth, amplitude fluctuation spectrum,
and intracavity Mandel Q-parameter. In the bad-cavity regime we
estimate the frequency at which the peak of relaxation
oscillations appears. With the help of the master equation for
density operator written in terms of coherent states we obtain
approximated expression for Glauber-Sudarshan P function. This
solution can be used in the case of good-cavity regime and allows
us to investigate in more detail the thresholdless behaviour of
the nanolaser. All two approaches are in a very good agreement
with each other and numerical simulations.
\end{abstract}
\date{\today}
\pacs{ }

\maketitle

\section{1. INTRODUCTION}

One of the first theoretical work devoted to possibility of
realization of a single-emitter laser was published by Mu and
Savage in~\cite{Mu1992}. Three- and four-level pumped emitters
placed into lossy cavity and interacted with cavity mode have been
considered there. Some interesting effects which do not appear in
conventional lasers have been revealed, such as squeezing,
self-quenching in incoherently pumped laser, deviations from the
Schawlow-Townes formula for the linewidth. Subsequent publications
related to single-emitter
problem~\cite{Ginzel1993,Pellizari1994,Loffler1997,Jones1999,Clemens2004,Kilin2001,Kilin2002,Kilin2012}
have proven these results and also discovered new effects
connected with the presence of one emitter: vacuum-Rabi doublet in
the spectrum~\cite{Ginzel1993,Loffler1997}, lasing without
inversion~\cite{Kilin2001}, entanglement between emitter and field
subsystems~\cite{Kilin2002}. In most of these works results have
been obtained with help of different type of numerical methods
applied to the master equation for density operator or to the
Heisenberg-Langevin equations.

Nowadays a single-emitter laser is realized in experiments where
the role of emitter can play the single atom~\cite{McKeever2003},
or ion~\cite{Dubin2010} or quantum dot~\cite{Nomura2009}. In this
connection, the purpose of our paper is to provide analytical
results which can be considered as a tools for experimenter. Thus
in our previous Brief Report~\cite{Larionov2011} with help of the
Heisenberg-Langevin approach we obtained analytical expressions
for linewidth, amplitude fluctuation spectrum and Mandel Q
parameter, which describe the behaviour of a single-emitter laser
in the case of good cavity regime. We also improved the condition
required for thresholdless regime. In this paper in Sec.~2 we will
consider and analyze in more detail the linearization procedure
which we used earlier and extend our calculations to bad-cavity
regime. In Sec.~3 we will write master equation for our nanolaser
in the coherent state representation and in stationary regime
derive approximated expression for Glauber-Sudarshan P function.
With the help of the latter we will prove our previous results,
analyze some limited cases and threshold behaviour of nanolaser.

\section{2. HEISENBERG-LANGEVIN\ APPROACH}

\subsection{2.1 Model. Semiclassical theory}

The simplest model of a single-emitter nanolaser is the single two-level
system placed inside the single-mode cavity and incoherently pumped to its
upper level. Just four constants characterize this nanolaser: $\Gamma $ --
the incoherent pumping rate, pumping process associated with the transition $%
\left\vert 1\right\rangle \rightarrow \left\vert 2\right\rangle $; $\gamma
/2 $ -- the decay rate of polarization due to spontaneous emission to modes
other than the laser mode; $\kappa /2$ -- the field decay rate in the
cavity; $g$ -- the coupling constant between the field and the two-level
system. The Heisenberg-Langevin equations for our nanolaser are%
\begin{eqnarray}
     \frac{d}{dt}\hat{a}(t)&=&-\frac{\kappa}{2}\hat{a}(t)
     +g\hat{\sigma}(t) +\hat{f}_a(t),
           \nonumber \\
     \frac{d}{dt}\hat{\sigma}(t)&=&-\frac{1}{2}(\Gamma+\gamma)\hat{\sigma}(t)
     +g\hat{D}(t)\hat{a}(t)+\hat{f}_{\sigma}(t),
           \nonumber \\
     \frac{d}{dt}\hat{D}(t)&=&\Gamma-\gamma-(\Gamma+\gamma)
     \hat{D}(t)
           \nonumber \\
     &-&2g\Bigl[\hat{\sigma}^{\dag}(t)\hat{a}(t)
     +\hat{a}^{\dag}(t)\hat{\sigma}(t)\Bigr]+\hat{f}_{D}(t).
           \label{H-L}
\end{eqnarray}
Here $\hat{a}(t)$ and $\hat{a}^{\dag }(t)$ are the photon annihilation and
creation operators in the cavity mode, $\hat{\sigma}=|1\rangle \langle 2|$
is the operator of polarization of the two-level system, and $\hat{D}%
=|2\rangle \langle 2|-|1\rangle \langle 1|$ is the operator of inversion
between the upper level $|2\rangle $ and the lower level $|1\rangle $ of the
system. We use the fact that for a single two-level system $|2\rangle
\langle 2|+|1\rangle \langle 1|=\mathbf{I}$ - where $\mathbf{I}$ is a unity
operator.

The $\hat{f}_{a}\left( t\right) $, $\hat{f}_{\sigma }\left( t\right) $ and $%
\hat{f}_{D}\left( t\right) $ are the Langevin noise operators which arise
through the interaction with the heat baths. Using the standard
quantum-optical methods (see, for example, Refs.~\cite{methods}), we obtain
the following nonzero correlation functions of these operators:
\begin{eqnarray}
     \langle\hat{f}_a(t)\hat{f}^{\dag}_a(t')\rangle &=& \kappa\delta(t-t'),
           \nonumber \\
     \langle\hat{f}^{\dag}_{\sigma}(t)\hat{f}_{\sigma}(t')\rangle &=& \Gamma\delta(t-t'),
           \nonumber \\
     \langle\hat{f}_{\sigma}(t)\hat{f}^{\dag}_{\sigma}(t')\rangle &=& \gamma\delta(t-t'),
           \label{Correlations} \\
     \langle\hat{f}_{\sigma}(t)\hat{f}_D(t')\rangle &=& 2\gamma\langle\hat{\sigma}(t)\rangle\delta(t-t'),
           \nonumber \\
     \langle\hat{f}_D(t)\hat{f}_{\sigma}(t')\rangle &=& -2\Gamma\langle\hat{\sigma}(t)\rangle\delta(t-t'),
           \nonumber \\
     \langle\hat{f}^{\dag}_{\sigma}(t)\hat{f}_D(t')\rangle &=&-2\Gamma\langle\hat{\sigma}^{\dag}(t)\rangle\delta(t-t'),
           \nonumber \\
     \langle\hat{f}_D(t)\hat{f}^{\dag}_{\sigma}(t')\rangle &=&2\gamma\langle\hat{\sigma}^{\dag}(t)\rangle\delta(t-t'),
           \nonumber \\
     \langle\hat{f}_D(t)\hat{f}_D(t')\rangle &=&2[(\Gamma+\gamma)-(\Gamma-\gamma)\langle\hat{D}(t)
     \rangle]\delta(t-t').
           \nonumber
\end{eqnarray}
We are also interested in the field properties transmitted outside
the cavity through the outcoupling mirror. For that we should
consider the annihilation operator $\hat{a}_{\rm{out}}(t)$ of
photons outside the
cavity~\cite{Gardiner2000}. The input-output transformation is%
\begin{equation}
     \hat{a}_{\rm out}(t)=\sqrt{\kappa}\hat{a}(t)-\hat{f}_a(t)/\sqrt{\kappa}.
        \label{Out}
\end{equation}

In the next Sec.~2.2 we will applied linearization procedure to
Eq.~(\ref{H-L}) which allows us to investigate the small
fluctuations of laser field near the dominant stationary
semiclassical mean value. To find the latter for our nanolaser we
need to collect a large number of photons in the cavity and also
provide the coherent interaction of each photon with the single
emitter.

In the assumption that above two conditions are satisfied let us
obtain the stationary semiclassical solution from Eq.~(\ref{H-L}).
For that we need to drop all time derivatives $d\hat{X}/dt$ and
average the remained stationary equations under assumption of following
decorrelations $\left\langle\hat{D}\hat{a}\right\rangle =%
\left\langle\hat{D}\right\rangle\left\langle\hat{a}\right\rangle$,
$\left\langle\hat{\sigma}^{\dag}\hat{a}\right\rangle%
=\left\langle\hat{\sigma}^{\dag}\right\rangle\left\langle\hat{a}%
\right\rangle $. Suppose that in steady state the optical phase is
randomly distributed between $0$ and $2\pi$ we take the average
value of operators with the fixed arbitrary mean value of the
phase to be equal to zero $\varphi _{0}=0$. Such selection of the
phase value implies the reality of semiclassical field amplitude
$a_{0}$ and polarization $\sigma _{0}=\frac{\kappa }{2g}a_{0}$.
Thus, the analytical expression for semiclassical stationary
intracavity intensity $I_{0}=|a_{0}|^{2}$ is therefore
\begin{equation}
     I_{0}=\frac{I_s}{2}\left[(r-1)-\frac{(r+1)^2}{c}\right].
        \label{SemiClass}
\end{equation}
where we introduce new dimensionless parameters: the dimensionless
pumping rate $r=\Gamma/\gamma$; the dimensionless saturation
intensity $I_{s}=\gamma/\kappa$; the dimensionless coupling
strength $c=4g^{2}/\kappa\gamma$.

Eq.~(\ref{SemiClass}) coincides with one firstly obtained by Mu
and Savage~\cite{Mu1992}. $I_{0}\left(r\right)$ is a parabolic
function of pump rate which has physical interpretation when $c>8$
and when the value of the pump rate lying in the domain between
two points, so called threshold $r_{th}$ and self-quenching
$r_{q}$ points, which are given by the following expressions%
\begin{eqnarray}
     r_{th}=r_{m}-\frac{c}{2}\sqrt{1-8/c},
     \text{ }r_{q}=r_{m}+\frac{c}{2}\sqrt{1-8/c},
     \label{r threshold}
\end{eqnarray}%
where $r_{m}=c/2-1$ is the point where the stationary solution has
a maximum $I_{m}=I_{s}\left(c/8-1\right)$.

The stationary intensity $I_{0}\left(r\right)$ increases as a pump
is increased and it starts from the threshold point $r_{th}$. When
$c\gg 8$ then $r_{th}\approx 1+4/c$, what indicates on finite
value of the threshold or, in other words, there is no
thresholdless regime in this semiclassical model. The
self-quenching point $r_{q}$ corresponds to pumping rate where the
atomic polarization is rapidly damped to zero due to emitter
trapping in the excited state, what leads to the damping of the
field.

From the expression for the maximum of intensity $I_{m}$ it follows that
large mean number of photons in the cavity can be achieved when $cI_{s}\gg 8$
(we assumed that $c\gg 8$). The latter condition can be written in other
form $g\gg \kappa $. Thus the semiclassical regime takes place when the
photons lifetime in the cavity is so long that each photon is provided by
coherent interaction with the single emitter (above mentioned condition).
The condition $cI_{s}\gg 8$ also gives us possibility to collect a large
mean number of photons in two opposite cases, namely in good - $I_{s}\gg 1$
and bad-cavity regimes $I_{s}\ll 1$.

\subsection{2.2 Linearization around the stationary solution}

To linearize the Heisenberg-Langevin equations (\ref{H-L}) we
assume that all operators can be presented as a sum of dominant
classical term $X_{0}$ and a "small" operator valued fluctuation
$\delta \hat{X}$ (see, for example, Refs.~\cite{Kolobov1993,
Haake1996}),
\begin{equation}
     \hat{X}=X_{0}+\delta\hat{X}.  \label{Sum}
\end{equation}%
After linearization with respect to $\delta\hat{X}$ we obtain the
following equations for the fluctuations
\begin{eqnarray}
     \frac{d}{dt}\delta\hat{a}(t)= &-&\frac{\kappa}{2}\delta\hat{a}(t)
     +g\delta\hat{\sigma}(t) +\hat{f}_{a}(t),
                 \nonumber \\
     \frac{d}{dt}\delta\hat{\sigma}(t)= &-&\frac{1}{2}(\Gamma+\gamma)\delta\hat{\sigma}(t)
     +g\Bigl[a_{0}\delta\hat{D}(t)+D_{0}\delta\hat{a}(t)\Bigr]
                  \nonumber \\
     &+&\hat{f}_{\sigma}(t),
                   \nonumber \\
     \frac{d}{dt}\delta\hat{D}(t)= &-&(\Gamma+\gamma)\delta\hat{D}(t)
     -2g\Bigl[\sigma _{0}(\delta\hat{a}(t)+\delta\hat{a}^{\dag}(t))
                     \nonumber \\
     &+&a_{0}(\delta\hat{\sigma}(t) +\delta\hat{\sigma}^{\dag }(t)\Bigr]+\hat{f}_{D}(t).
                      \label{H-L Linearized}
\end{eqnarray}

Before solving Eqs.~(\ref{H-L Linearized}) we first split
operators into Hermitian "real" and "imaginary" parts as%
\begin{eqnarray}
      \delta\hat{a}(t)&=&\delta\hat{u}(t)+i\delta\hat{\nu}(t),
               \nonumber \\
      \delta\hat{\sigma}(t)&=&\delta\hat{\mu}(t)+i\delta\hat{\eta}(t),
                \nonumber \\
      \hat{f}(t)&=&\hat{\Sigma}(t)+i\hat{\Delta}(t).
                 \label{Hermitian Oper}
\end{eqnarray}
In this way the linearized equations of motion separate into two
independent blocks%
\begin{eqnarray}
     \frac{d}{dt}\delta\hat{u}(t)=&-&\frac{\kappa}{2}\delta\hat{u}(t)
     +g\delta\hat{\mu}(t)+\hat{\Sigma}_{a}(t),
                  \nonumber \\
     \frac{d}{dt}\delta\hat{\mu}(t)=&-&\frac{1}{2}(\Gamma+\gamma)\delta\hat{\mu}(t)
     +g\Bigl[a_{0}\delta\hat{D}(t)+D_{0}\delta\hat{u}(t)\Bigr]
                   \nonumber \\
     &+&\hat{\Sigma}_{\sigma}(t),
                   \nonumber \\
     \frac{d}{dt}\delta\hat{D}(t)=&-&(\Gamma+\gamma)\delta\hat{D}(t)
     -4g\Bigl[\sigma _{0}\delta \hat{u}(t)+a_{0}\delta\hat{\mu}(t)\Bigr]
                    \nonumber \\
     &+&\hat{f}_{D}(t);
     \label{Re Block}
\end{eqnarray}%
\begin{eqnarray}
     \frac{d}{dt}\delta\hat{\nu}(t)&=&-\frac{\kappa}{2}\delta\hat{\nu}(t)
     +g\delta\hat{\eta}(t)+\hat{\Delta}_{a}(t),
               \nonumber \\
     \frac{d}{dt}\delta\hat{\eta}(t)&=&-\frac{1}{2}(\Gamma+\gamma)\delta\hat{\eta}(t)
     +gD_{0}\delta\hat{\nu}(t)+\hat{\Delta}_{\sigma}(t).
                 \nonumber \\
      \label{Im Block}
\end{eqnarray}%
We want to note that the independence of "real" and "imaginary"
parts is a result of our phase selection in the derivation of
semiclassical solution (see Eq.~(\ref{SemiClass}) and text above).

The block for three real parts $\delta\hat{u}$, $\delta\hat{\mu}$,
$\delta\hat{D}$ are related to the intensity fluctuations via
$\delta\hat{I}=2a_{0}\delta \hat{u}$. The two imaginary parts
$\delta\hat{\nu}$, $\delta\hat{\eta}$ can be associated with phase
fluctuations.

Eqs.~(\ref{Re Block}, \ref{Im Block}) can be resolved by means of Fourier transformation.
For that we need to perform the fluctuations as%
\begin{equation}
     \delta\hat{X}(t)=\int_{-\infty}^{\infty}\delta\hat{X}(\Omega)
     \exp(-i\Omega t)\frac{d\Omega}{2\pi},
     \label{Fourier}
\end{equation}%
which gives us the linear algebraic equations which can be resolved by means
of Cramer's rule. The result for the Fourier-transformed amplitude $\delta\hat{u}$
and phase $\delta\hat{\nu}$ quadrature components are%
\begin{widetext}
\begin{eqnarray}
     \delta\hat{u}(\Omega)
     &=&\frac{\hat{\Sigma}_{a}(\Omega)A(\Omega)+\hat{\Sigma}_{\sigma}(\Omega)C(\Omega)
     +\hat{f}_{D}(\Omega)B}{i\Omega\Bigl[(i\Omega-(\Gamma+\gamma+\kappa))/2)
     (i\Omega-(\Gamma+\gamma))+4g^{2}|a_{0}|^{2}\Bigr]-4\kappa g^{2}|a_{0}|^{2}},
        \label{Fourier du} \\
     \delta\hat{\nu}(\Omega)&=&\frac{g\hat{\Delta}_{\sigma}(\Omega)
     -\hat{\Delta}_{a}(\Omega)(i\Omega-(\Gamma+\gamma)/2)}
     {i\Omega\Bigl[i\Omega-(\Gamma+\gamma+\kappa)/2\Bigr]},
        \label{Fourier dv}
\end{eqnarray}%
\end{widetext}
where%
\begin{eqnarray}
     A(\Omega)&=&-[(i\Omega-(\Gamma+\gamma)/2)(i\Omega-(\Gamma+\gamma))
     +4g^{2}|a_{0}|^{2}] ,
         \nonumber \\
     C(\Omega)&=&g(i\Omega-(\Gamma+\gamma)),
     \text{ }B=-g^{2}a_{0}.
     \label{Const 1}
\end{eqnarray}
As it follows from the correlation function definition
Eqs.~(\ref{Correlations}) and from Eq.~(\ref{Fourier}) the
Fourier-transformed quadrature components $\delta\hat{u}$,
$\delta\hat{\nu}$ are $\delta$-function correlated. Using the
input-output transformation Eq.~(\ref{Out}) we find the amplitude
and phase fluctuation spectra%
\begin{widetext}
\begin{eqnarray}
     \langle\delta\hat{u}_{\rm{out}}(\Omega)\delta\hat{u}_{\rm{out}}
     (\Omega^{\prime})\rangle=\frac{1}{4}\delta(\Omega+\Omega^{\prime})
     \Bigl[1+\frac{\kappa^{2}}{4}\frac{a\Omega^{2}+b\gamma^{2}}
     {\Omega^{2}(\Omega^{2}-d)^{2}+\gamma^{2}(e\Omega^{2}-f)^{2}}\Bigr],
         \label{Amplitude Spectrum} \\
     \langle\delta\hat{\nu}_{\rm{out}}(\Omega)\delta\hat{\nu}_{\rm{out}}
     (\Omega^{\prime})\rangle=\frac{1}{4}\delta(\Omega+\Omega^{\prime})
     \Bigl[1+\frac{\kappa^{2}}{4}\frac{\gamma^{2}[c(r+1)+(r+1)^{2}]}
     {\Omega^{2}(\Omega^{2}+\gamma^{2}(r+1+1/I_{s})^{2}/4)}\Bigr],
        \label{Phase Spectrum}
\end{eqnarray}%
\end{widetext} where
\begin{eqnarray}
     a&=&\gamma^{2}[c(r-1)-(r+1)^{2}+2cr],
          \nonumber \\
     b&=&\gamma^{2}(r-1)[3(r+1)^{3}+2c^{2}-c(r+1)(r-5)],
          \nonumber \\
     d&=&\gamma^{2}\frac{cI_{s}(r-1)+(r+1)}{2I_{s}},\text{ } e=\frac{3(r+1)+1/I_{s}}{2},
          \nonumber \\
     f&=&\gamma^{2}cI_{0}/I_{s}^{2}.
     \label{Const 2}
\end{eqnarray}%
Now we will use these results to analyze two different regimes of
our nanolaser, namely good- and bad-cavity regimes.

\subsection{2.3 Good-cavity regime}

In the case of good-cavity regime $I_{s}\gg 1$ the obtained
expression for the amplitude fluctuation spectrum
Eq.~(\ref{Amplitude Spectrum}) becomes
\begin{equation}
     \langle\delta\hat{u}_{\rm{out}}(\Omega)\delta\hat{u}_{\rm{out}}
     (\Omega^{\prime})\rangle=\frac{1}{4}\delta(\Omega+\Omega^{\prime})
     \Bigl[1+\frac{S(r,c)}{1+(\Omega/\Omega_{0})^{2}}\Bigr].
       \label{Amplitude Spectrum GC}
\end{equation}%
The first term in the square brackets corresponds to standard quantum limit
(SQL), the second term is a Lorentzian with width $\Omega _{0}$ given by
\begin{equation}
     \Omega _{0}=\kappa\Bigl[1-\frac{(r+1)^{2}}{c(r-1)}\Bigr]
       \label{Width}
\end{equation}%
and this term describes the additionally exes above the SQL or
reduction of the noise below SQL depending of the sign of the
strength $S(r,c)$
\begin{equation}
     S(r,c)=\frac{(r-1)\left[3(r+1)^{3}+2c^{2}-c(r+1)(r-5)\right]}
     {\left[c(r-1)-(r+1)^{2}\right]^{2}}.
   \label{Strength S}
\end{equation}

We have investigated this expression in order to find out a
possibility of the noise reduction below the SQL in the laser
light outside the cavity. Unfortunately, this kind of nonclassical
phenomenon are very limited. We have found that in the region of
pump parameter $r_{m}<r<r_{q}$ the strength $S(r,c)$ is always
positive and increasing function, resulting in the excess noise in
Eq.~(\ref{Amplitude Spectrum GC}). The small negative values of
$S(r,c)$ are observed when the pump parameter is centered around
$r=c/5$ ($c\geq 200$), i.e. lies in the region $r_{th}\ll
r<r_{m}$. Probably, these negative values are the result of the
well-known antibunching phenomenon for a single-emitter. However,
this antibunching effect is strongly attenuated due to the effect
of accumulation of many photons inside the cavity during the long
(and random) photon lifetime.

The Mandel Q-parameter can be expressed through the amplitude
quadrature component $\delta\hat{u}$ as
$Q=4\cdot\int_{-\infty}^{\infty}\langle\delta \hat{u}(\Omega)
\delta\hat{u}(-\Omega)\rangle d\Omega/2\pi-1$ and the result is%
\begin{equation}
     Q(r,c)=\frac{1}{2}S(r,c)\left[1-\frac{(r+1)^{2}}{c(r-1)}\right].
     \label{Mandel Q}
\end{equation}%
We see that the strength $S(r,c)$ is also characterized this
quantity. On the Fig.~1a there are some graphics for Mandel
Q-parameter for different coupling constant $c$ when $I_{s}\gg 1$.
For the pump rate lying between threshold and quenching points our
theoretical results are in a very good agreement with the exact
solution of the master equation for the density matrix (see next
Sec.~3). The main divergency appears near the threshold and
quenching points where the linear theory does not work. Two
fluctuations peaks of Mandel Q-parameter are well
known~\cite{Jones1999}: first of them associated with laser turn
on and second broad peak associated with laser turn off.

Another quantity of interest is the low-frequency asymptotic
version of the fluctuation spectrum of the phase quadrature
component. Using the Eq.~(\ref{Phase Spectrum}) we obtain
\begin{equation}
     \langle\delta\hat{\nu}_{\rm{out}}(\Omega)
     \delta\hat{\nu}_{\rm{out}}(\Omega^{\prime})\rangle
     =\delta(\Omega+\Omega^{\prime})\frac{\kappa^{2}}{4}\frac{1+c/(r+1)}{\Omega^{2}}.
      \label{Phase Spectrum GC}
\end{equation}%
The low-frequency divergence in this spectrum as $1/\Omega^{2}$ is
manifesting for the phase diffusion process~\cite{Courtois1991}.
This spectrum also defines the linewidth $\Delta\nu$ of our
nanolaser through
$\langle(\delta\hat{\nu}_{\rm{out}})^{2}\rangle_{\Omega}/I_{0}$
\begin{equation}
     \Delta\nu=\Delta\nu_{\rm{ST}}\frac{1}{2}\left(1+\frac{c}{r+1}\right),
       \label{Linewidth}
\end{equation}%
where $\Delta\nu_{\rm{ST}}=\kappa/(2I_{0})$ is the Schawlow-Townes
linewidth of a conventional incoherently pumped laser. The
obtained formula for the linewidth shows the deviation from the
Schawlow-Townes result and this coincides with the numerical
simulations performed by Clemens et al~\cite{Clemens2004}.

At the end of this section we want to discuss the legality of
above linearization procedure. Let us explain it in terms of
adiabatic elimination of the emitter variables in Eq.~(\ref{H-L}).
Indeed, in the case when $I_{s}\gg 1$ one can obtain the
Heisenberg-Langevin equation only for the field operator
$\hat{a}(t)$. The semiclassical solution of the latter coincides
with Eq.~(\ref{SemiClass}). The following linearization with
respect to fluctuation $\delta\hat{a}(t)$ gives us equation which
contains the products like $\delta\hat{a}(t)\hat{f}_{\alpha}(t)$
($\alpha=\sigma$, $D$). Because of presence just one emitter they
can not be neglected in frame of linearization procedure. However,
these products appear in the equation as terms multiplied by small
factor $2\kappa/(\Gamma-\gamma)<1/I_{0}\ll 1$, what makes the
order of the products similar to second order in fluctuation
$\delta\hat{a}(t)$. Finally, the fully linearized equation for
$\hat{a}(t)$ gives the same results as was obtained in this
section.
\begin{figure}
\centerline{\includegraphics[width=7cm]{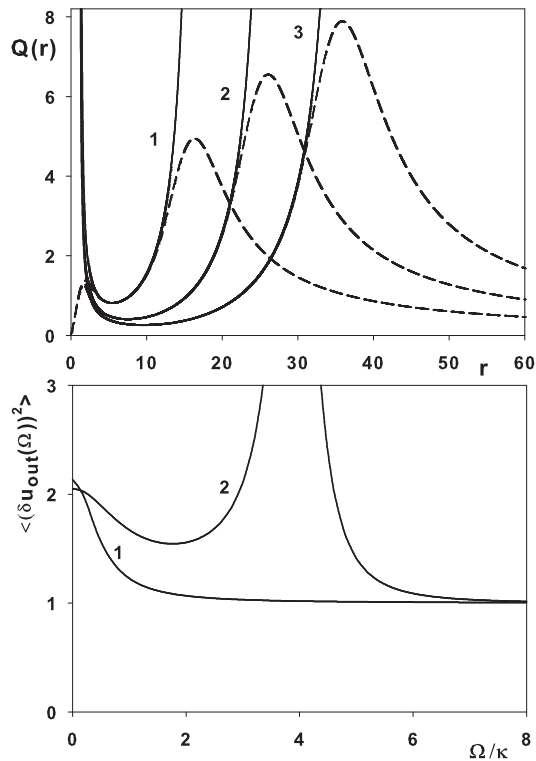}} \caption{a) The
Mandel Q-parameter (\protect\ref{Mandel Q}) vs the pump parameter
$r$. Dashed line shows results obtained with help of function
$P_{0}$ (\protect\ref{P0}). 1) $c=20$; 2) $c=30$; 3) $c=40$. For
all curves $I_{s}=20$. b) The spectrum of amplitude fluctuations.
1) $r=r_{m}=199$, 2) $r=3$. For all curves $I_{s}=0.2$,
$c=400$.\label{fig1}}
\end{figure}
\subsection{2.4 Bad-cavity regime}

The considered single-emitter laser relates to so called high
$\beta$ lasers - the lasers with high fraction $\beta$ of
spontaneous emission into the lasing mode (see expression for
$\beta$ in Sec.~3, Eq.~(\ref{beta})). Most of present work on the
intensity noise in high $\beta$ lasers has been applied to
standard semiconductor laser diodes, embedding quantum wells as
gain material. These models predict that if $\beta$ is small, the
peak-to-valley difference of the relaxation oscillations is very
large and it decreases with increasing $\beta$~\cite{Bjork1994,
Protsenko2001}. However, all these models are based on standard
rate equations used in quantum well lasers and very few studies
have been carried out in this field when switching from quantum
well lasers to quantum dot lasers. Some recent research was
focused on the turn-on dynamics and relaxation oscillation in
standard quantum dots lasers~\cite{Ludge2008} but has not been
extended to unconventional high $\beta$ nanolasers.

Here we use results obtained in Sec.~2.2 to analyze behaviour of
our nanolaser in the case of bad-cavity regime $I_{s}\ll 1$. As
was mentioned above, for that we only need to preserve the
following condition $cI_{s}\gg 8$.

If the pump rate $r$ is close to the maximum point $r_{m}$ the
amplitude fluctuations spectrum is described by the same
Eqs.~(\ref{Amplitude Spectrum GC}, \ref{Phase Spectrum GC}) as in
a good-cavity regime (see Fig.~1b, curve 1). The main difference
is the magnitude of the width $\Omega_{0}$ (\ref{Width}) for the
amplitude fluctuations spectrum.

Other situation takes place when the pump rate $r$ lies not far
from the threshold point $r_{th}$. In this case the high peak in
the amplitude fluctuations spectrum appears (see Fig.~1b, curve
2). The physical origin of this peak is the relaxation oscillation
phenomenon. Because of linearization procedure break down in the
vicinity of threshold $r_{th}$, we can not correctly define width
and height of this peak, but we can estimate the frequency of the
relaxation oscillations $\Omega_{osc}/\kappa
=I_{s}\sqrt{c(r-1)/2}$.

\section{3. MASTER EQUATION APPROACH}

This section are devoted to investigation of the nanolaser
properties with help of master equation for density matrix. This
approach have been used by many authors. Thus in~\cite{Mu1992} the
master equation have been reduced to system of first order,
ordinary differential equations and analyzed numerically.
In~\cite{Jones1999, Clemens2004} more efficient quantum trajectory
algorithm have been developed for numerical evaluating of master
equation. The works~\cite{Kilin2001,Kilin2002} are devoted to
analysis of the master equation written in terms of coherent
states for field. In light of our interest, we want to briefly
discuss one of them, namely~\cite{Kilin2001}. Authors have worked
with the system of equations for Glauber-Sudarshan P function and
additional quasi-probabilities. Besides numerical simulations they
have also found approximated expression for P function which can
be used in the case of good-cavity regime. Here we follow the
nearest approach as in~\cite{Kilin2001}, but we manage to derive
isolated stationary equation for P function. A detailed analysis
of the latter allows us to obtain approximated solution, which has
wider region of application than that in~\cite{Kilin2001}.

\subsection{3.1 Coherent state representation of master equation}

The master equation for our nanolaser is (see, for
example,~\cite{Clemens2004})
\begin{eqnarray}
     \frac{\partial\hat{\rho}}{\partial t}=
     &-&\frac{i}{\hbar}\Bigl[\hat{V},\hat{\rho}\Bigr]
     +\frac{\kappa}{2}\left(2\text{ }\hat{a}\hat{\rho}\hat{a}^{\dag}
     -\hat{a}^{\dag}\hat{a}\hat{\rho}
     -\hat{\rho}\hat{a}^{\dag}\hat{a}\right)
           \nonumber \\
     &+&\frac{\gamma}{2}\left(2\hat{\sigma}\hat{\rho}\hat{\sigma}^{\dag}
     -\hat{\sigma}^{\dag}\hat{\sigma}\hat{\rho}
     -\hat{\rho}\hat{\sigma}^{\dag }\hat{\sigma}\right)
             \nonumber \\
     &+&\frac{\Gamma}{2}\left(2\hat{\sigma}^{\dag }\hat{\rho}\hat{\sigma}
     -\hat{\sigma}\hat{\sigma}^{\dag}\hat{\rho}
     -\hat{\rho}\hat{\sigma}\hat{\sigma}^{\dag}\right),
           \label{Master Eq}
\end{eqnarray}%
where interaction between the cavity mode and the single two-level
system is given by the Jaynes-Cummings Hamiltonian
\begin{equation}
     \hat{V}=i\hbar
     g\left(\hat{a}^{\dag}\hat{\sigma}-\hat{\sigma}^{\dag}\hat{a}\right).
            \label{J-C Hamiltonian}
\end{equation}
Finally, we want to obtain equation for P function, for that we
need to rewrite Eq.~(\ref{Master Eq}) in terms of coherent states
for field $|z\rangle $, $|z^{\ast}\rangle $ and in projections of
the density operator on the two-level system states $|1\rangle $
and $|2\rangle $. Using well known rules (see, for example, Ref.~\cite{RulesCoherent})%
\begin{eqnarray}
     \hat{a}\hat{\rho} &\rightarrow& z\hat{\rho}\left(z,z^{\ast}\right),
           \nonumber \\
     \hat{\rho}\hat{a} &\rightarrow& \left[z-\frac{\partial}{\partial z^{\ast}}\right]
     \hat{\rho}\left(z,z^{\ast}\right),
            \nonumber \\
     \hat{\rho}\hat{a}^{\dag} &\rightarrow& z^{\ast}\hat{\rho}\left(z,z^{\ast}\right),
             \nonumber \\
     \hat{a}^{\dag}\hat{\rho} &\rightarrow& \left[z^{\ast}-\frac{\partial}
     {\partial z}\right]\hat{\rho}\left(z,z^{\ast}\right),
          \label{RulesCoherent}
\end{eqnarray}%
where $z$, $z^{\ast}$ are the complex variables and introduce the
following quasi-probabilities $\rho _{ik}\left(z,z^{\ast}\right)
=\left\langle i\right\vert \hat{\rho}\left(z,z^{\ast}\right)
\left\vert k\right\rangle \equiv \rho _{ik}$ ($i,k=1,2$),
$D=\rho_{22}\left(z,z^{\ast}\right)-\rho_{11}\left(z,z^{\ast}\right)$
and P function
$P=\rho_{11}\left(z,z^{\ast}\right)+\rho_{22}\left(z,z^{\ast}\right)$,
we obtain from Eq.~(\ref{Master Eq}) the system of partial
differential equations
\begin{eqnarray}
     \frac{\partial P}{\partial t}&=&\frac{\partial}{\partial z}
     \left(\frac{\kappa}{2}zP-g\rho_{21}\right)
     +\frac{\partial}{\partial z^{\ast}}
     \left(\frac{\kappa}{2}z^{\ast}P-g\rho _{12}\right),
           \nonumber \\
     \frac{\partial D}{\partial t}&=&(\Gamma-\gamma)P-(\Gamma+\gamma)D
     +\frac{\partial}{\partial z}\left(\frac{\kappa }{2}zD
     +g\rho_{21}\right)
             \nonumber \\
     &+&\frac{\partial}{\partial z^{\ast}}
     \left(\frac{\kappa}{2}z^{\ast}D+g\rho_{12}\right)
     -2g\left[z^{\ast}\rho_{21}+z\rho_{12}\right],
              \nonumber \\
     \frac{\partial \rho_{21}}{\partial t}&=&-\frac{\left(\Gamma+\gamma\right)}{2}\rho_{21}
     +\frac{\partial}{\partial z}\left(\frac{\kappa}{2}z\rho_{21}\right)
     +\frac{\partial}{\partial z^{\ast}}\left(\frac{\kappa}{2}z^{\ast}\rho_{21}\right)
               \nonumber \\
     &+&g\left[zD-\frac{1}{2}\frac{\partial}{\partial z^{\ast}}\left(P+D\right)\right].
                \label{Master Eq Coh Repr}
\end{eqnarray}
The first equation in this system is written in special form which
can be associated with conservation law for the quasi-probability
$\partial P/\partial t=div\overline{J}$, where $\overline{J}$ is
known as probability current density. The additional
quasi-probabilities $D$ and $\rho_{21}$ have a clear physical
meaning. Thus, the mean values for inversion and polarization of
the two-level system are expressed through $D$ and $\rho_{21}$ as
$\left\langle \hat{D}\right\rangle =\int
D\left(z,z^{\ast}\right)d^{2}z$,
$\left\langle\hat{\sigma}\right\rangle =\int
\rho_{21}\left(z,z^{\ast}\right)d^{2}z$.

\subsection{3.2 Stationary solution}

In stationary regime the probability current density
$\overline{J}$ can be equated with zero (see, for
example,~\cite{Mandel,Risken}) what gives the following relations
$\rho_{21}=\left(\rho_{12}\right)^{\ast}=\left(\kappa/2g\right)zP$.
The latter allows us to eliminate $\rho_{21}$, $\rho_{12}$ from
the stationary form of the system (\ref{Master Eq Coh Repr}) and
obtain two coupled equations for $P$ and $D$. In the new variables
$I=\left\vert z\right\vert^{2}$ and $\varphi$
($z=\sqrt{I}\exp\left( i\varphi\right)$) these equations read
\begin{eqnarray}
     IP-\frac{I_{s}}{2}\left[\left(r-1\right)P-\left(r+1\right)D\right]
     =\frac{1}{2}\frac{\partial}{\partial I}I\left(P+D\right),
           \label{Master Eq Steady 1} \\
     IP\frac{\left(r+1+1/I_{s}\right)}{c}
     -\left[ID+\frac{1}{2}\left(P+D\right)\right]
         \nonumber \\
     =\frac{1}{2}\frac{\partial}{\partial I}\left[\frac{4}{cI_{s}}I^{2}P-I\left(P+D\right)\right]
     -\frac{i}{4}\frac{\partial}{\partial \varphi}\left(P+D\right),
            \label{Master Eq Steady 2}
\end{eqnarray}%
where we introduced above mentioned dimensionless parameters
$I_{s}$, $c$, $r$.

There is no phasing influence on our nanolaser, e.g. - external
field with fixed phase or coherent pumping. Therefore expected
stationary solution does not depend on phase and we can write $P$
and $D$ as only function of $I=\left\vert z\right\vert^{2}$. This
is also following from the second Eq.~(\ref{Master Eq Steady 2}):
the term proportional to imaginary unit should be equated to zero
because of reality of functions $P$ and $D$.

Let us derive isolated equation for the P function in the
stationary regime. Using Eqs.~(\ref{Master Eq Steady 1},
\ref{Master Eq Steady 2}) we can express function $D$ in terms of
$P$
\begin{widetext}
\begin{equation}
     D=\frac{1}{(r+1-1/I_{s})-2I/I_{s}}\left(\frac{4}{cI_{s}^{2}}I^{2}
     \frac{\partial}{\partial I}P+\frac{2\left[3-I_{s}\left(r+1+c\right)\right]}
     {cI_{s}^{2}}IP+(r-1+1/I_{s})P\right).
     \label{D function}
\end{equation}
\end{widetext} Now put this relation into the first or the second
equation in the system Eqs.~(\ref{Master Eq Steady 1}, \ref{Master
Eq Steady 2}). In such a way we get isolated equation for $P$
function
\begin{eqnarray}
     &P^{\prime\prime}\left(I\right)+p(I)P^{\prime}\left(I\right)+q(I)P\left(I\right)=0,
        \nonumber \\
     &p(I)=\left(a_{10}+a_{11}I+a_{12}I^{2}\right)/\left(a_{02}I^{2}+a_{03}I^{3}\right),
         \nonumber \\
     &q(I)=\left(a_{20}+a_{21}I+a_{22}I^{2}\right)/\left(a_{02}I^{2}+a_{03}I^{3}\right),
         \label{P Steady Eq}
\end{eqnarray}%
where the prime indicates derivative with respect to variable
$I=\left\vert z\right\vert^{2}$ and functions $p(I)$, $q(I)$
depend on physical parameters $I_{s}$, $c$, $r$ via $a_{ik}$ (see
Appendix).

The obtained equation is the second order differential equation
with polynomial coefficients. To use this equation we need to
define boundary $P(0)$, $P(\infty)$ or "initial" $P\left(0\right)
$, $P^{\prime}(0)$ conditions. If we equate the variable $I$ to
zero in Eq.~(\ref{P Steady Eq}) we obtain the following relation
$P^{\prime}(0)=-a_{20}/a_{10}P\left(0\right)$. This relation
together with normalization of the quasi-probability $P$ gives a
comfortable way to define the "initial" condition for numerical
simulations.

In the case of good-cavity regime functions $p(I)$ and $q(I)$ in
Eq.~(\ref{P Steady Eq}) contain large parameters like $cI_{s}$,
$I_{s}^{3}$ or $I_{s}^{2}$ (see Appendix). Therefore, we can try
to find approximated solution of Eq.~(\ref{P Steady Eq}) using the
perturbation method (see, for example, Ref.~\cite{Nayfeh}). The
point of this method is representation of the unknown function $P$
as iterative series $P=\sum_{n=0}^{\infty}\lambda^{-n}P_{n}$,
where $\lambda$ is a large parameter. A substitution of the latter
series into Eq.~(\ref{P Steady Eq}) with subsequent equating to
zero of terms with same powers $1/\lambda$ gives possibility to
find the different orders of approximations $P_{n}$.

To isolate the correct single large parameter let us rewrite
$p(I)$ and $q(I)$ in the following form
\begin{eqnarray}
     p(I)=a_{12}\left(I-I_{-1}\right)\left(I-I_{+1}\right)/I^{2}\left(I-I_{00}\right),
       \label{pol p new} \\
     q(I)=a_{22}\left(I-I_{-2}\right)\left(I-I_{+2}\right)/I^{2}\left(I-I_{00}\right),
       \label{pol q new}
\end{eqnarray}%
where roots are
\begin{equation}
     I_{\pm i}=-a_{i1}/2a_{i2}\pm\sqrt{\left(a_{i1}/a_{i2}\right)^{2}-4a_{i0}/a_{i2}}/2,
          \nonumber \\
     \text{ }I_{00}=-a_{02}.
     \nonumber
\end{equation}%
In the conditions $I_{s}\gg 1$, $c>8$ the magnitudes of the roots
$I_{+1}$, $I_{+2}$, $I_{00}$ approximately equal each other and
the roots $I_{-1}$, $I_{-2}$ have the similar order. So, it is
natural to save all $I_{\pm i}$ and extract the large parameter
from the constants $a_{12}$, $a_{22}$ in Eqs.~(\ref{pol p new},
\ref{pol q new}). This parameter is therefore $\lambda
=cI_{s}\left(=a_{22}\approx -a_{12}\right)$ and Eq.~(\ref{P Steady
Eq}) can be written as follows
\begin{equation}
     \lambda^{-1}P^{\prime\prime}\left(I\right)+\tilde{p}(I)P^{\prime}\left(I\right)
     +\tilde{q}(I)P\left(I\right)=0,
         \label{P Steady Eq Lambda}
\end{equation}%
where $\tilde{p}(I)=p(I)/a_{22}$, $\tilde{q}(I)=q(I)/a_{22}$. Here
we want to note that we do not neglect the term $r/c$ in relation
$a_{12}/a_{22}=\left(7-3I_{s}(r+1)-2cI_{s}\right)/2cI_{s}$,
because of pump rate $r$ can be comparable or bigger than $c$ (for
example near to maximum $r_{m}$ or to quenching $r_{q}$ points).

Thus, in zero-order approximation (first term in the series
$P=\sum_{n=0}^{\infty}\lambda^{-n}P_{n}$) we neglect the term in
Eq.~(\ref{P Steady Eq Lambda}) proportional to $1/\lambda$ what
implies first-order differential equation for function $P_{0}$
\begin{eqnarray}
     &a_{12}\left(I-I_{-1}\right)\left(I-I_{+1}\right)P_{0}^{\prime}\left(I\right)&
       \nonumber \\
     +&a_{22}\left(I-I_{-2}\right)\left(I-I_{+2}\right)P_{0}\left(I\right)=0.&
        \label{P0 Steady Eq}
\end{eqnarray}%
$P_{0}$ is then found from Eq.~(\ref{P0 Steady Eq}):
\begin{widetext}
\begin{eqnarray}
     P_{0}\left(I\right)=\left\{
     \begin{array}{cc}
          N_{0}\left(1-I/I_{-1}\right)^{f_{1}}\left(1-I/I_{+1}\right)^{f_{2}}
          \exp\left(-\frac{a_{22}}{a_{12}}I\right),
          &\text{if}\text{ }I\leq I_{-1} \\
          0,
          &\text{otherwise}
     \end{array}\right.\,
           \label{P0} \\
     f_{1}=-\frac{a_{22}}{a_{12}}\frac{\left(I_{-1}-I_{-2}\right)
     \left(I_{-1}-I_{+2}\right)}{\left(I_{-1}-I_{+1}\right)}, \text{ }
     f_{2}=\frac{a_{22}}{a_{12}}\frac{\left(I_{+1}-I_{-2}\right)
     \left(I_{+1}-I_{+2}\right)}{\left(I_{-1}-I_{+1}\right)},
          \nonumber
\end{eqnarray}
\end{widetext}%
where $N_{0}$ is normalizing constant. The roots and consequently
the powers $f_{1}$, $f_{2}$ in Eq.~(\ref{P0}) have complex
structure and will be estimated below in some special cases.

To understand why our solution is restricted we note that the
roots $I_{\pm 1}$ of function $p(I)$ in Eq.~(\ref{P Steady Eq})
are known as turning points, where oscillating behaviour of
$P\left(I\right)$ can change on exponential. A careful examination
of Eq.~(\ref{P Steady Eq}) shows that in the conditions $I_{s}\gg
1$, $c>8$ on the $I$-domain ranging from $0$ to turning point
$I_{-1}$ the $P\left(I\right)$ is positive and nonoscillating
function. Moreover, its essential changing occurs just on latter
domain (it is not for every set of parameters $r$, $I_{s}$, $c$,
see below). Otherwise, when $I>I_{-1}$ then $P\left(I\right)$
becomes oscillating function. The obtained solution Eq.~(\ref{P0})
can not describe any oscillations, moreover it possesses imaginary
values when $I>I_{-1}$, therefore we have restricted definition
domain for our solution by the value $I_{-1}$ and call them
boundary root.

To reveal physical meaning of the boundary root we have simplified
expression for $I_{-1}$ and found maximum of function
$P_{0}\left(I\right)$ on the domain $\left(0,I_{-1}\right)$. When
$I_{s}\gg 1$, $c>8$ we have managed to estimate the latter
quantities with help of results obtained in Sec.~2.3:
$I_{-1}\approx I_{s}r/\left(2+3r/c\right)\geq
I_{0}\left[1+Q\left(r,c\right)\right]$ and the point where
$P_{0}\left(I\right)$ reaches his maximum is $I=I_{-2}\approx
I_{0}$. Thus, P function has a maximum at the point corresponding
to the semiclassical intensity $I_{0}$\ Eq.~(\ref{SemiClass}) and
the boundary root is at a distance $I_{0}Q\left(r,c\right)$ from
it, where $Q\left(r,c\right)$ is the Mandel Q-parameter
Eq.~(\ref{Mandel Q}).

In the Fig.~2 we plot the P function calculated using
Eq.~(\ref{P0}) (dashed line) and using numerical simulations of
the equation (\ref{P Steady Eq}) (solid gray line). The different
curves correspond to different values of pump rate $r$ while $c$
and $I_{s}$ are fixed. We see a very good agreement with numerical
results.

The main divergency appears for value of pump rate discovered in
Sec.~2.3, namely for $r=c/5$, where the Mandel Q-parameter
$Q\left(r,c\right)$ (\ref{Mandel Q}) has a minimum (see
Eq.~(\ref{Strength S}) and text below). In the right wing of P
function (Fig.~2b) the small oscillations appear when $c$ becomes
comparable with $I_{s}$, but still our solution is good. With
growing of $c$ the oscillations become more pronounced and the
maximum of P function goes out of domain $\left(0,I_{-1}\right)$
and our solution (\ref{P0}) becomes inapplicable in vicinity of
$r=c/5$. It is interesting to note that critical situation occurs
when $c\geqslant 200$. For latter values of $c$ and for arbitrary
saturation intensity $I_{s}$ the maximum of P function goes out of
the domain $\left(0,I_{-1}\right)$. As mentioned above, the P
function acquires oscillating character for $I>I_{-1}$, what
probably indicates on the nonclassical effect associated with
photon antibunching.

Now let us analyze the obtained solution (\ref{P0}) in some
special cases.
\begin{figure}
\centerline{\includegraphics[width=7cm]{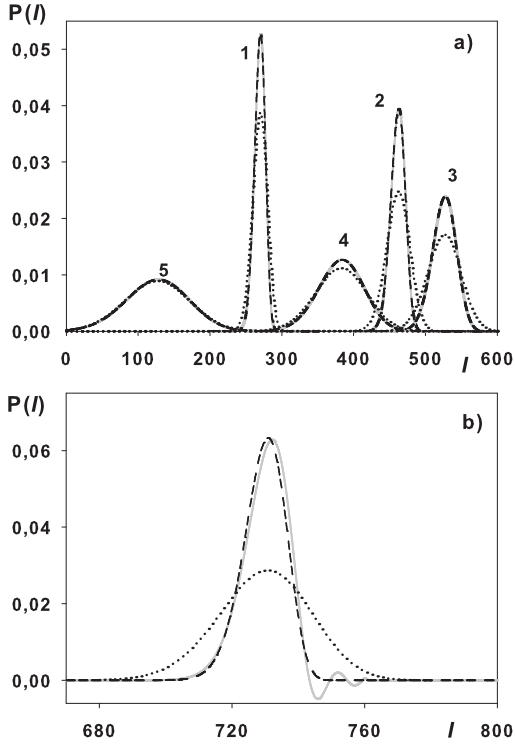}} \caption{a) P
function vs $I=\left\vert z\right\vert ^{2}$ for different pump
parameter $r$. 1) $r=8$, 2) $r=16$, 3) $r=24$, 4) $r=36$, 5)
$r=44$. For all curves $I_{s}=100$, $c=50$; b) Oscillation
behaviour of the P function, $I_{s}=c=100$, $r=c/5=20$. Dashed
line - solution (\ref{P0}), solid gray line - numerical
simulation, dotted line - solution from
Ref.~\cite{Kilin2001}.\label{fig2}}
\end{figure}
\subsubsection{Far below threshold and far above quenching}

At first we consider situation when the pump rate lies far below
threshold $r\ll r_{th}$. In this case the average value of
$I=\left\vert z\right\vert^{2}$ is small and the following
inequalities $I/I_{-1}\approx I/\left(I_{s}r/2\right)\ll 1$,
$I/I_{+1}\approx I/(I_{s}\left( r+1\right) /2)\ll 1$ are
satisfied. Thus, the factor in front of exponential in
Eq.~(\ref{P0}) can be approximated as follows
\begin{eqnarray}
     &\left(1-I/I_{-1}\right)^{f_{1}}\left(1-I/I_{+1}\right)^{f_{2}}
         \nonumber \\
     &=\exp\left[f_{1}\ln \left(1-I/I_{-1}\right)\right]
     \exp\left[f_{2}\ln \left(1-I/I_{+1}\right)\right]
          \nonumber \\
     &\approx \exp \left[-\left(f_{1}/I_{-1}+f_{2}/I_{+1}\right)I\right].
           \label{Factor Thermal}
\end{eqnarray}%
The $P_{0}$ is therefore
\begin{equation}
     P_{0}\left(I\right)=\frac{1}{\left(a_{10}/a_{20}\right)}
     \exp\left(-\frac{I}{\left(a_{10}/a_{20}\right)}\right).
            \label{Thermal Dist}
\end{equation}%
The obtained expression for P function corresponds to thermal
distribution with the following mean number of photons in the
cavity $\left\langle \hat{n}\right\rangle=a_{10}/a_{20}\sim r$.

For the values of pump rate lying far above quenching point $r\gg
r_{q}$ we can make the same manipulation with Eq.~(\ref{P0}) as
above. Using inequalities $I/I_{-1}\approx
I/(I_{s}r/\left(2+3r/c\right))\ll 1$, $I/I_{+1}\approx
I/(I_{s}r/2)\ll 1$ which occur in this case, we get the same
thermal distribution as Eq.~(\ref{Thermal Dist}), but the mean
number of photons possesses another behaviour $\left\langle
\hat{n}\right\rangle=a_{10}/a_{20}\sim 1/r$.

In the two limiting situations, when $r\rightarrow 0$ or
$r\rightarrow \infty$, the corresponding mean number of photons
converge to zero and the distribution Eq.~(\ref{Thermal Dist})
gives Dirac delta-function
$P_{0}\left(I\right)=\delta\left(I\right)$, what coincides with
the thermal distribution with zero temperature.

\subsubsection{Near to maximum point $r_{m}$}

Here we consider situation when the pump rate is in the vicinity
of $r_{m}$, where the semiclassical solution $I_{0}$
Eq.~(\ref{SemiClass}) reaches his maximum. In this case the powers
in Eq.~(\ref{P0}) are simplified as $f_{1}\approx
I_{s}\left(3\left(r+1\right)^{3}+2c^{2}-c\left(r+1\right)\left(r-5\right)\right)
/\left(3\left(r+1\right)+2c\right)^{2}$, $f_{2}\approx 0$ and the
root $I_{-2}$ becomes similar to $I_{0}$. Simple algebra leads to
following expression for $P_{0}$
\begin{eqnarray}
     P_{0}\left(\Delta I\right)=\left\{
     \begin{array}{cc}
          N_{0}\left[\left(1-\Delta I/\Lambda\right)^{\Lambda}
          \exp\left(\Delta I\right)\right]^{\alpha},
          &\text{if}\text{ }\Delta I\leqslant\Lambda \\
          0,
          &\text{otherwise}
     \end{array}\right.\
          \nonumber \\
          \label{P0 1}
\end{eqnarray}%
where we introduced new variable $\Delta I=I-I_{0}$ and constants
$\Lambda=f_{1}/\alpha$,
$\alpha=2c/\left(3\left(r+1\right)+2c\right)$. For chosen values
of pump rate the average value of $\Delta I$ is small and the
inequality $\Delta I/\Lambda\ll 1$ is satisfied. Thus for the
factor in front of exponential in Eq.~(\ref{P0 1}) we can write%
\begin{eqnarray}
     \left(1-\Delta I/\Lambda\right)^{\Lambda}
     &=&\exp\left[\Lambda\ln \left(1-\Delta I/\Lambda\right)\right]
        \nonumber \\
     &\approx& \exp\left[-\Delta I-\Delta I^{2}/2\Lambda \right],
         \label{Factor G}
\end{eqnarray}%
what implies Gaussian P function%
\begin{equation}
     P_{0}\left(I\right)=N_{0}\exp\left[-\frac{\left(I-I_{0}\right)^{2}}
     {2\Lambda /\alpha}\right].
         \label{Gaussian Dist}
\end{equation}%
Here we removed the restriction on $P_{0}$. Indeed, the
distribution (\ref{Gaussian Dist}) never really sees the boundary
and $I$ can be taken to run from $0$ to $\infty$.

Using the obtained Eq.~(\ref{Gaussian Dist}) we can estimate the
average value of photons in the cavity
$\left\langle\hat{n}\right\rangle$ and its variance
$\langle\left(\Delta \hat{n}\right)^{2}\rangle$%
\begin{eqnarray}
     \left\langle\hat{n}\right\rangle
     &=&\int_{0}^{\infty}IP_{0}\left(I\right)dI\approx I_{0},
       \label{n Gauss} \\
     \langle\left(\Delta\hat{n}\right)^{2}\rangle
     &=&\left\langle\hat{n}\right\rangle+\int_{0}^{\infty}
     \left(I-\left\langle\hat{n}\right\rangle\right)^{2}P_{0}\left(I\right)dI
         \nonumber \\
     &\approx& I_{0}\left[1+\Lambda/\left(\alpha I_{0}\right)\right].
        \label{var Gauss}
\end{eqnarray}%
A substitution of the explicit form of $\Lambda$ and $\alpha$ into
the expression for variance gives
$\langle\left(\Delta\hat{n}\right)^{2}\rangle=I_{0}\left[1+Q\left(r,c\right)\right]$,
where $Q\left(r,c\right)$ is the Mandel Q-parameter obtained from
linear theory (\ref{Mandel Q}).

In the weak coupling regime $c<I_{s}$ the obtained
Eq.~(\ref{Gaussian Dist}) works good for values of pump rate lying
in whole semiclassical region $r_{th}<r<r_{q}$. In the strong
coupling regime $c>I_{s}$ the symmetrical Gaussian distribution
(\ref{Gaussian Dist}) is not sufficient and it becomes more
pronounced near to mentioned value of pump rate $r=c/5$.

\subsubsection{Threshold}

In our previous report~\cite{Larionov2011} we considered nanolaser
behaviour around the semiclassical threshold $r_{th}$ (\ref{r
threshold}) and specified the condition required for thresholdless
regime. Here we continue our research using the obtained
Eq.~(\ref{P0}).

The behaviour of P function in the threshold point $r_{th}$ for
three different regimes 1) $c\ll I_{s}$, 2) $c\approx I_{s}$, 3)
$c\gg I_{s}$ is shown in Fig.~3a. To realize these regimes we fix
the large coupling constant $c=100$ and set the different values
of the saturation intensity: $I_{s}=600$, $60$, $6$ (we choose
such values of $I_{s}$ for better resolution of P function
behaviour). In the weak coupling regime ($c\ll I_{s} $) the P
function has a typical plateau (it is marked by solid straight
line, curve 1), which indicates transition to lasing: from thermal
to Gaussian type distribution. As saturation intensity is
decreased and the strong coupling regime ($c\gg I_{s}$) occurs the
maximum of P function moves from zero value of variable $I$ and
the semiclassical threshold behaviour disappears (curve 3).

Above we obtained the maximum point for $P_{0}$ (\ref{P0}) and it
was the root $I_{-2}$. To define the threshold value of pump rate
we need to solve equation $I_{-2}\left(r\right)=0$, i.e. find such
value of $r$ when the maximum of the P function is in the point
$I=I_{-2}\left(r\right)=0$. The approximate solution is
$\tilde{r}=1+4/c-2/I_{s}$. When $c\ll I_{s}$ then the last term
can be neglected and $\tilde{r}$ equal to semiclassical threshold
$r_{th}\approx 1+4/c>1$ (\ref{r threshold}). When $c\gg I_{s}$
then $\tilde{r}$ becomes smaller than unity what indicates the
disappearance of the semiclassical threshold and transition to
thresholdless regime (see the vanishing of narrow peak in the
behaviour of Mandel Q parameter in our previews
report~\cite{Larionov2011}, Fig.~2). The dynamics of $\tilde{r}$
in above three considered regimes is $\tilde{r}\approx 1.04$, $1$,
$0.66$ (remind that $r_{th}=1.04$).

At the end of this section we want to correct the misprint
occurred in our previous report (see formula (13)
in~\cite{Larionov2011}). The valid expression for fraction $\beta$
of spontaneous emission into the laser mode is
\begin{equation}
     \beta=c/\left[\left(c+1\right)+I_{s}\left(r+1\right)\right].
         \label{beta}
\end{equation}%
We should note that this misprint did not affect on all our results and
discussions.
\begin{figure}
\centerline{\includegraphics[width=7cm]{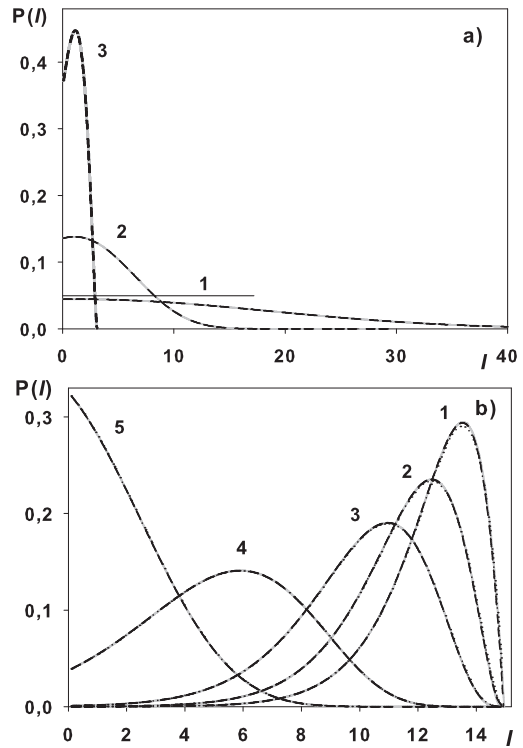}} \caption{P
function vs $I=\left\vert z\right\vert^{2}$. a) Transition to the
thresholdless regime. 1) $I_{s}=600$, 2) $I_{s}=60$, 3) $I_{s}=6$.
For all curves $c=100$, $r=r_{th}=1.04$. b) Parameters was taken
as in Ref.~\cite{Kilin2001}: 1) $c=2000$, $I_{s}=5$, $r=6$, 2)
$c=1428.57$, $I_{s}=7$, $r=4.28$, 3) $c=1000$, $I_{s}=10$, $r=3$,
4) $c=500$, $I_{s}=20$, $r=1.5$, 5) $c=285.7$, $I_{s}=35$,
$r=0.86$. Dashed line - solution (\ref{P0}), solid gray line -
numerical simulation, dotted line - solution from
Ref.~\cite{Kilin2001}.\label{fig3}}
\end{figure}
\subsubsection{Comparing with other authors}

In introduction to this section we mentioned that close approach
based on quasi-probabilities have been considered by two authors
Karlovich and Kilin in~\cite{Kilin2001}. They have worked with
first order differential equation system for function P and
additional quasi-probabilities. By neglecting of small parameters
in this system they have obtained integrable equations and as a
result the analytical formula for P function. The structure of
their solution has an identical form with our Eq.~(\ref{P0}) (see
formula (9) in~\cite{Kilin2001}), but this solution works good in
the region near to semiclassical threshold $r_{th}$ or quenching
points $r_{q}$. For the pump rate lying around the maximum point
$r_{m}$ it gives uncorrect result for width of function $P$ and as
a consequence the uncorrect variance of the photon distribution.

In the Fig.~3b we plot some graphics of quasi-probability P for
parameters taken from~\cite{Kilin2001}. It is clear to see that
our Eq.~(\ref{P0}) gives the same result as a solution (9)
obtained in~\cite{Kilin2001} and all of them are in a good
agreement with numerical simulations. The Fig.~2a,b demonstrate
that for the pump rate lying around the maximum point $r_{m}$ our
Eq.~(\ref{P0}) (dashed line) has an advantage with
(9)~\cite{Kilin2001} (dotted line). The latter indicates well the
maximum of true P function, but the width is uncorrect.

The derivation of a single equation for P function (\ref{P Steady Eq}) and
accurate extraction of large parameters gives us possibility to obtained the
solution which has more wide region of application.

\section{4. SUMMARY}

In this paper we have studied physical properties of a
single-emitter laser. The problem have been investigated in terms
of the Heisenberg-Langevin equations and in terms of the master
equation for density matrix. In both approaches we have provided
analytical results which are summarized below.

In the case of good-cavity regime with help of the
Heisenberg-Langevin approach we have obtained analytical
expressions for linewidth Eq.~(\ref{Linewidth}), amplitude
fluctuation spectrum Eq.~(\ref{Amplitude Spectrum GC}) and Mandel
Q parameter Eq.~(\ref{Mandel Q}). These results work good in the
semiclassical region of pumping rate $r_{th}<r<r_{q}$. According
to nanolaser behaviour the latter region can be split into two
subregions. In the first subregion $r<r_{m}$ the nanolaser
behaviour is similar to that of conventional laser. Also in this
subregion for $r$ centered around $r=c/5$ and for large coupling
constant ($c\geqslant 200$) we have discovered the small noise
reduction below the SQL in the laser light outside the cavity and
small negative values in $Q\left(r,c\right)$ (\ref{Mandel Q}). The
master equation approach confirms this nonclassical phenomenon: P
function manifests oscillating behaviour. Probably, these negative
values are the result of the well-known antibunching phenomenon
for a single-emitter. In the second subregion $r>r_{m}$ the excess
noise is always takes place. The P function in this subregion does
not have any nonclassical features.

In the bad-cavity regime we have observed two different
situations: when the pump rate $r$ is close to its maximum value
$r_{m}$ then laser generates as in the good-cavity regime; when
$r$ is close to threshold $r_{th}$ then high peak in the amplitude
fluctuations spectrum appears which indicates on the relaxation
oscillation phenomenon.

With the help of the master equation for density matrix written in
terms of coherent states we have managed to derive the stationary
equation for the Glauber-Sudarshan P function~(\ref{P Steady Eq}).
A detailed analysis of this equation allowed us to obtain
approximated solution Eq.~(\ref{P0}), which works good when
$cI_{s}\gg 1$. We have analyzed Eq.~(\ref{P0}) in some special
cases. In the good-cavity regime in the semiclassical region
$r_{th}<r<r_{q}$ Eq.~(\ref{P0}) can be written as Gaussian
distribution with mean value coincides with the semiclassical
intracavity intensity $I_{0}$ (\ref{SemiClass}) and with width
$I_{0}Q\left(r,c\right)$ (\ref{Mandel Q}). For the values of pump
rate lying far below threshold and far above self-quenching points
Eq.~(\ref{P0}) corresponds to thermal distribution.

Using Eq.~(\ref{P0}) we have also obtained approximated expression
for threshold pump rate $\tilde{r}=1+4/c-2/I_{s}$. In the
weak-coupling regime ($I_{s}\gg c$) the last term can be neglected
what implies the semiclassical threshold $\tilde{r}=r_{th}\approx
1+4/c$. When strong-coupling regime ($I_{s}\ll c$) occurs then
$\tilde{r}<r_{th}$ what indicates on the transition to
thresholdless regime.

\section{Acknowledgments}

This work was supported by French National Agency (ANR) through
Nanoscience and Nanotechnology Program (Project NATIF
n$^{\circ}$ANR-09-NANO-012-01), by the CNRS-RFBR collaboration
(CNRS 6054 and RFBR 12-02-91056) and by external fellowship of the
Russian Quantum Center (Ref. number 86).

\section{Appendix}

The constants $a_{ik}$ in Eq.~(\ref{P Steady Eq})%
\begin{eqnarray}
     a_{02}&=&\frac{1}{2}-I_{s}\frac{\left(r+1\right)}{2},
         \nonumber \\
     a_{03}&=&1,
         \nonumber \\
     a_{10}&=&I_{s}^{2}\frac{cr}{4}-I_{s}^{3}\frac{cr(r+1)}{4},
         \nonumber \\
     a_{11}&=&\frac{9}{4}-I_{s}\frac{6(r+1)+c}{2}+I_{s}^{2}\frac{3(r+1)^{2}+c(4r+2)}{4},
         \nonumber \\
     a_{12}&=&\frac{7}{2}-I_{s}\frac{3(r+1)+2c}{2},
         \nonumber \\
     a_{20}&=&\frac{3}{2}-I_{s}\frac{11(r+1)+3c}{4}+I_{s}^{2}\frac{3(r+1)^{2}+2c}{2}
          \nonumber \\
     &+&I_{s}^{3}\frac{c(r+1)}{4}\left[(r-1)-\frac{(r+1)^{2}}{c}\right],
         \nonumber \\
     a_{21}&=&\frac{3}{2}-2I_{s}(r+1)-I_{s}^{2}\frac{2cr-(r+1)^{2}}{2},
         \nonumber \\
     a_{22}&=&cI_{s}.
              \label{a(ik)}
\end{eqnarray}

\end{document}